\documentclass [twocolumn,showpacs,preprintnumbers,amsmath,amssymb,aps]{revtex4}
\usepackage {graphicx}
\usepackage {dcolumn}
\usepackage {bm}

\begin {document}


\title {Parameters of the Effective Singlet-Triplet Model for Band 
Structure of High-$T_c$ Cuprates by Different Approaches}

\author {M.M. Korshunov}
\email {mkor@iph.krasn.ru}
\author {V.A. Gavrichkov}%
\author {S.G. Ovchinnikov}
\affiliation {%
L.V. Kirensky Institute of Physics, Siberian Branch of Russian Academy 
of Sciences, 660036 Krasnoyarsk, Russia
}%

\author {Z.V. Pchelkina}
\author {I.A. Nekrasov}
\author {M.A. Korotin}
\author {V.I. Anisimov}
\affiliation {
Institute of Metal Physics, Ural Branch of Russian Academy of Sciences, 
620219 Ekaterinburg GSP-170, Russia}%

\date {\today}

\begin {abstract}

The present paper covers the problem of parameters determination for
High-$T_c$ superconductive copper oxides. Different approaches, {\it ab
initio} LDA and LDA+U calculations and Generalized Tight-Binding (GTB)
method for strongly correlated electron systems, are used to calculate
hopping and exchange parameters of the effective singlet-triplet model for
$CuO_2$-layer. The resulting parameters are in remarkably good agreement
with each other and with parameters extracted from experiment. 
This set of parameters is proposed for proper quantitative description 
of physics of hole doped High-$T_c$ cuprates in the framework of
effective models.

\end {abstract}

\maketitle

\section {Introduction}

High-$T_c$ superconductive cuprates (HTSC) belong to the class of 
substances where the strong electron correlations are important. This 
circumstance and also the fact that these substances have non-trivial 
phase diagrams (see e.g. review \cite {qqq1}) lead to the difficulties  in
description of HTSC in framework of first principles ({\it ab initio}) 
methods, especially in the low doping region. So at present moment the most
adequate method of HTSC theoretical investigations is the model approach.
Effective models of HTSC (e.g. $t-J$ model) usually contain free parameters
that could be fitted to experimental data (comparison of calculated and
experimental Fermi surfaces, dispersion curves, etc.) but as soon as one
uses model approach the question concerning correctness of these parameters
arises. One of the possible ways to answer this question is to obtain
relations between parameters of some effective model and microscopic
parameters of the underlying crystal structure. The underlying crystal
structure of HTSC could be described either by the 3-band Emery model \cite
{Emery,Varma} or by the multiband $p-d$ model \cite {Gaididei}. One can
compare parameters bound to these models with parameters obtained by very
different approach, e.g. with {\it ab initio} calculated parameters. It
does not mean that the {\it ab initio} band structure is correct. Due to
the strong electron correlations it is for sure incorrect in  the low
doping region where these correlations are most significant. Nevertheless, 
the single electron parameters are of interest and may be compared with the
appropriate parameters obtained by fitting to the experimental ARPES data.

In the present paper we obtain relations between microscopic parameters of
the multiband $p-d$ model and parameters of the effective singlet-triplet
$t-J$ model for hole doped HTSC. Then we compare these parameters and the
$t-J$ model parameters obtained in the {\it ab initio} calculations. 
In Section \ref {section:abinitio} the details of {\it ab initio} calculation
within density functional theory are presented. In Section \ref {section:Heff}
formulation of the effective singlet-triplet model as the
low-energy Hamiltonian for the multiband $p-d$ model with generalized
tight-binding (GTB) method applied are described. In both methods the parent insulating
compound $La_2CuO_4$ is investigated. Parameters are obtained at zero
doping because within the GTB method the
evolution of the band structure with doping is described only by changes in
the occupation numbers of zero-hole, single-hole, and two-hole local terms,
while all parameters are fitted in undoped case and therefore fixed for all
doping levels. The resulting parameters of both approaches (GTB and {\it ab
initio}) are in very good qualitative and quantitative agreement with each
other and parameters extracted from experiment. Also, these parameters are in
reasonable agreement with $t-J$ model parameters used in the literature. We
conclude that the obtained set of model parameters should be used in
effective models for proper quantitative description of HTSC in the whole
doping region.

\section {{\it Ab initio} Parameters Calculation 
\label {section:abinitio}}

The band structure of $La_2CuO_4$ was obtained in frames of the linear
muffin-tin orbital  method \cite {lmto} in tight-binding approach \cite
{tblmto} (TB-LMTO) withing the local density  approximation (LDA). The
crystal structure data \cite {struc} corresponds to tetragonal $La_2CuO_4$.
The effective hopping parameters $t_{\rho}$ were calculated  by the least
square fit procedure to the bands obtained in LDA calculation \cite {Sorella}. The
effective exchange interaction parameters $J_{\rho}$ were calculated  using
the formula derived in \cite {exch}, where the Green function method was
used to calculate $J_{\rho}$ as second derivative of the ground state
energy with respect to the magnetic moment rotation  angle via eigenvalues
and eigenfunctions obtained in LDA+$U$ calculation \cite {lda_u}. The
LDA+$U$ approach allows to obtain the experimental antiferromagnetic insulating ground 
state for the undoped cuprate in contrast to the LDA approach which gives a
nonmagnetic metallic ground state \cite {lda_u}. 
The Coulomb parameters $U=10$~eV and $J=1$~eV used in LDA+$U$
calculation  were obtained in constrained LSDA supercell calculations \cite
{constrain}.

\section {GTB method and formulation of the effective singlet-triplet model
\label {section:Heff}}

The $t-J$ \cite {Chao} and Hubbard \cite {Hubbard} models are widely used
to investigate  HTSC compounds. While using these models one, in principle,
can catch up qualitatively  essential physics. The parameters in these
models ({\it i.e.} the hopping integral $t$, antifferomagnetic exchange
$J$, Hubbard repulsion $U$) are typically extracted from experimental data.
Thus, these parameters do not have a direct microscopical meaning. More
consequent  approach is to write down the multiband Hamiltonian for the
real crystal structure (which now  includes parameters of this {\it real}
structure) and map this Hamiltonian onto some low-energy model (like $t-J$
model). In this case parameters of real structure could be taken from the
{\it ab initio} calculations or fitted to experimental data.

As the starting model that properly describes crystal structure of the
cuprates it  is convenient to use 3-band Emery $p-d$ model \cite
{Emery,Varma} or the multiband $p-d$  model \cite {Gaididei}. The set of
microscopic parameters for the first one was calculated in
\cite {Hybertsen,McMahan}. While this model is simplier then the multiband
$p-d$  model it lacks for some significant features, namely importance of
$d_{z^2}$ orbitals on copper and $p_z$ orbitals on apical oxygen. Non-zero
occupancy of $d_{z^2}$ orbitals pointed out in XAS and EELS experiments
which shows 2-10\% occupancy of $d_{z^2}$ orbitals \cite {Bianconi,Romberg}
and 15\% doping dependent occupancy of $p_z$ orbitals \cite {Chen} in all
HTSC of p-type (hole doped). In order to take into account
these facts the multi-band $p-d$ model should be used:
\begin {eqnarray}
H_ {pd} &=&\sum\limits_ {f, \lambda, \sigma}(\epsilon_ {\lambda}-\mu) n_ {f
\lambda \sigma} + \sum\limits_ {<f, g>} \sum\limits_ {\lambda, \lambda',
\sigma} T_ {f g}^ {\lambda \lambda'} c_ {f \lambda \sigma}^ {+} c_ {g
\lambda' \sigma} \nonumber \\ &+& \frac {1}{2}\sum\limits_ {f, g, \lambda,
\lambda'} \sum\limits_ {\sigma_ 1, \sigma_ 2, \sigma_ 3, \sigma_ 4} V_ {f
g}^ {\lambda \lambda'} c_ {f \lambda \sigma_1}^ {+} c_ {f \lambda \sigma_
3} c_ {g \lambda' \sigma_ 2}^ {+} c_ {g \lambda' \sigma_ 4},
\label {eq:Hpd}
\end {eqnarray}
where $c_{f \lambda \sigma}$ is the annihilation operator in Wannier
representation of the  hole at site $f$ (copper or oxygen) at orbital
$\lambda$ with spin $\sigma$,  $n_{f \lambda \sigma}=c_{f \lambda
\sigma}^{+} c_{f \lambda \sigma}$. Indexs $\lambda$  run through
$d_{x^2-y^2} \equiv d_x$ and $d_{3z^2-r^2} \equiv d_z$ orbitals on copper, 
$p_x$ and $p_y$ atomic orbitals on the plane oxygen sites and $p_z$ orbital
on the apical  oxygen; $\epsilon_{\lambda}$ - single-electron energy of the
atomic orbital $\lambda$.  $T_{f g}^{\lambda \lambda'}$ includes matrix
elements of hoppings between copper and  oxygen ($t_{pd}$ for hopping $d_x
\leftrightarrow p_x,p_y$; $t_{pd}/\sqrt{3}$ for  $d_z \leftrightarrow
p_x,p_y$; $t'_{pd}$ for $d_z \leftrightarrow p_z$) and between  oxygen and
oxygen ($t_{pp}$ for hopping $p_x \leftrightarrow p_y$; $t'_{pp}$ for 
hopping $p_x,p_y \leftrightarrow p_z$). The Coulomb matrix elements  $V_{f
g}^{\lambda \lambda'}$ includes intraatomic Hubbard repulsions of two
holes  with opposite spins on one copper and oxygen orbital ($U_d$, $U_p$),
between different  orbitals of copper and oxygen ($V_d$, $V_p$), Hund
exchange on copper and oxygen  ($J_d$, $J_p$) and the nearest-neighbor
copper-oxygen Coulomb repulsion $V_{pd}$.

GTB method \cite {Ovchinnikov} consist of exact diagonalization of
intracell part of $p-d$  Hamiltonian (\ref {eq:Hpd}) and perturbative account
for the intercell part.  For $La_{2-x}Sr_xCuO_4$ the unit cell is $CuO_{6}$
cluster, and a problem of  nonorthogonality of the molecular orbitals of
adjacent cells is solved by  an explicit fashion namely by constructing the
relevant Wannier functions on a  five-orbitals initial basis of the atomic
states \cite {Gavrichkov1,Gavrichkov2}.  In a new symmetric basis the
intracell part of the total Hamiltonian is diagonalized,  allowing to
classify all possible effective quasiparticle excitations in
$CuO_{2}$-plane  according to a symmetry.

\begin {figure}
\includegraphics[width=\linewidth]{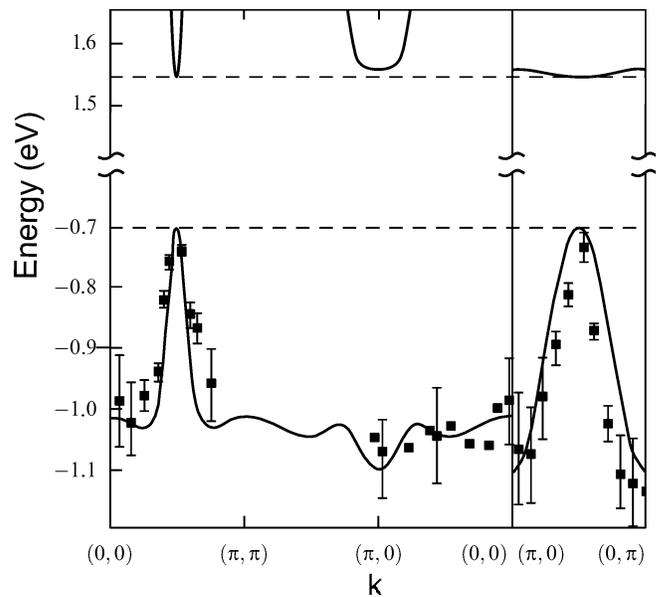}
\caption {\label{fig1} GTB method dispersion (doping concentration $x=0$) of the top 
of the valence band and the bottom of the conduction band divided by the insulating 
gap is shown. Horizontal dashed lines marks the in-gap states which spectral weight 
is proportional to $x$. Points with error bars represents experimental ARPES data 
for $Sr_2CuO_2Cl_2$ \cite {Wells}.}
\end {figure}

Calculations \cite {Gavrichkov1,Gavrichkov2} of the quasiparticle
dispersion and spectral  intensities in the framework of multiband $p-d$
model with use of GTB method are in very  good agreement with ARPES data on
insulating compound $Sr_2CuO_2Cl_2$ \cite {Wells,Durr}  (see Fig.~\ref
{fig1}).

\begin {figure}
\includegraphics[width=\linewidth]{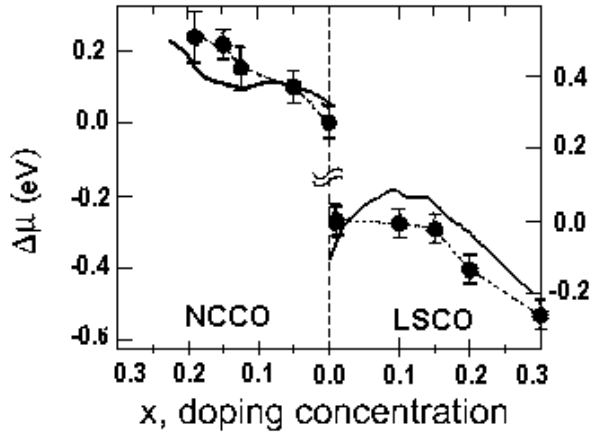}
\caption {\label {fig2} Dependence of chemical potential shift $\Delta\mu$ on 
concentration of doping $x$ for $Nd_{2-x}Sr_xCuO_4$ and $La_{2-x}Sr_xCuO_4$. 
Straight lines are results of GTB calculations, filled 
circles with error bars are experimental points \cite {Harima}.}
\end {figure}

Other significant results of this method are \cite {Borisov1,Borisov2}:

i) Pinning of Fermi level in $La_{2-x}Sr_xCuO_4$ at low concentrations was
obtained  in agreement with experiments \cite {Ino1,Harima}. This pinning
appears due to the in-gap state,  spectral weight of this state is
proportional to doping concentration $x$ and when Fermi  level comes to
this in-gap band then Fermi level ``stacks'' there. In Fig.~\ref {fig2} the 
doping dependence of chemical potential shift $\Delta\mu$ for n-type 
High-$T_c$ $Nd_{2-x}Sr_xCuO_4$ (NCCO) and p-type High-$T_c$
$La_{2-x}Sr_xCuO_4$ (LSCO)  is shown. The localized in-gap state exist in
NCCO also for the same reason as in LSCO,  but its energy is determined by
the extremum of the band at $(\pi/2,\pi/2)$ point and it  appears to be
above the bottom of the conductivity band. Thus, the first doped electron 
goes into the band state at the $(\pi,0)$ and the chemical potential for
the very small  concentration merges into the band. At higher $x$ it meets
the in-gap state with a pinning  at $0.08<x<0.18$ and then $\mu$ again
moves into the band. The dependence $\mu(x)$ for NCCO  is quite
asymmetrical to the LSCO and also agrees with experimental data
\cite {Harima}.

ii) Experimentally observed \cite {Ino2} evolution of Fermi surface with
doping from hole-type (centered at $(\pi,\pi)$) in the underdoped region to
electron-type (centered at $(0,0)$) in the overdoped region is
qualitatively reproduced in this method.

iii) Pseudogap feature for $La_{2-x}Sr_xCuO_4$ is obtained as a lowering of
density of states between the in-gap state and the states at the top of the
valence band.

Above results was obtained with the following set of the microscopic parameters:
\begin {eqnarray} \label {eq:pparams}
\begin {array}{llll}
\varepsilon_ {d_ {x^2-y^2}}=0, & \varepsilon_ {d_ {z^2}}=2, & \varepsilon_
{p_x}=1.5, & \varepsilon_ {p_z}=0.45, \\ t_ {pd}=1, & t_ {pp}=0.46, & t'_
{pd}=0.58, & t'_ {pp}=0.42,\\ U_ d=V_ d=9, & J_ d=1, J_ p=0, & U_ p=V_ p=4,
& V_ {pd}=1.5.
\end {array}
\end {eqnarray}

As the next step we will formulate the effective model. Simplest way to do
it is to completely neglect contribution of two-particle triplet state
$^3B_{1g}$. Then there will be only one low-energy two-particle state -
Zhang-Rice-type singlet $^1A_{1g}$ - and the effective model will be the
usual $t-J$ model. But in the multiband $p-d$ model the difference
$\epsilon_T-\epsilon_S$ between energy of two-particle singlet and
two-particle triplet depends strongly on various model parameters,
particularly on distance of the apical oxygen from the planar oxygen,
energy of the apical oxygen, difference between energy of
$d_{z^2}$-orbitals and $d_{x^2-y^2}$-orbitals. For the realistic values of
model parameters $\varepsilon_T-\varepsilon_S$ is close to $0.5$~eV
\cite {Gavrichkov2,Raimondi} contrary to the 3-band model with this value
being about $2$~eV (this case was considered in \cite {Zaanen,Hayn}). To
take into account triplet states we will derive the effective Hamiltonian
for multiband $p-d$ model by exclusion of the intersubband hopping between
low (LHB) and upper (UHB) Hubbard subbands, similar to \cite {Chao}.

As the Hubbard X-operators $X_{f}^{p\,q} \equiv \left|
p\right\rangle\left\langle q\right|$ on site $f$ represents natural
language to describe strongly correlated electron systems in the rest of
the paper we will use these operators. The X-operators are constructed in
the Hilbert space that consists of a vacuum $n_{h}=0$ state  $| 0 \rangle$,
single-hole $| \sigma \rangle = \left\{ | \uparrow \rangle, | \downarrow
\rangle \right\}$ state of $b_{1g}$ symmetry, two-hole singlet state $| S
\rangle$ of $^1A_{1g}$ symmetry and two-hole triplet state $| TM \rangle$
(where $M=+1,0,-1$) of $^3B_{1g}$ symmetry.

We write the Hamiltonian in the form $H=H_0+H_1$, where the excitations via
the charge transfer gap $E_{ct}$ are included in $H_1$. Then we define an
operator $H(\epsilon)=H_0+\epsilon H_1$ and make the unitary transformation
$\tilde{H}(\epsilon) = exp{\left(-i \epsilon \hat{S} \right)} H
\left(\epsilon\right) exp{\left( i \epsilon \hat{S} \right)}$. Vanishing
linear in $\epsilon$ component of $ \tilde{H}(\epsilon) $ gives the
equation for matrix $\hat{S}$: $H_1 + i \left[ H_0 ,\hat{S} \right]=0$. The
effective Hamiltonian is obtained in second order in $\epsilon$ and at
$\epsilon=1$ is given by:
\begin {equation}
\tilde {H}=H_ 0+\frac {1}{2} i \left[ H_ 1, \hat {S} \right].
\label {eq4}
\end {equation}
Thus, for the multiband $p-d$ model (\ref {eq:Hpd}) in case of electron doping
(n-type systems) we obtain the usual $t-J$ model:
\begin {eqnarray}
H_ {t-J}&=&\sum_ {f, \sigma}\varepsilon_ 1 X_ {f}^ {\sigma \sigma} + \sum_
{<f, g>, \sigma} t_ {fg}^ {0 0} X_ {f}^ {\sigma 0} X_ {g}^ {0 \sigma}
\nonumber \\ &+& \sum_ {<f, g>} J_ {fg} \left( \vec {S}_ f \vec {S}_ g -
\frac {1}{4} n_ f n_ g \right), 
\label {eq:HtJ}
\end {eqnarray}
where $\vec{S}_f$ are spin operators and $n_f$ are number of particles
operators. The $J_{fg} = 2 \left( t_{fg}^{0S} \right)^{2} / E_{ct}$ is the
exchange integral, $E_{ct}$ is the energy of charge-transfer gap (similar
to $U$ in the Hubbard model, $E_{ct} \approx 2$~eV for cuprates). Chemical
potential $\mu$ is included in $\varepsilon_1$.

For p-type systems effective Hamiltonian has the form of a singlet-triplet $t-J$
model \cite {Korshunov}:
\begin {equation}
H=H_0 + H_t + \sum_{<f,g>}J_{fg} \left( \vec{S}_f \vec{S}_g - \frac{1}{4} n_f n_g \right), \label{eq:Heff}
\end {equation}
where $H_0$ (unperturbated part of the Hamiltonian) and $H_t$
(kinetic part of $H$) are given by the expressions:
\begin {eqnarray*}
H_0 &=& \sum_{f} \left[ \varepsilon_{1} \sum_{\sigma} X_{f}^{\sigma \sigma}
+ \varepsilon_{2S} X_{f}^{S S} + \varepsilon_{2T} \sum_{M} X_{f}^{TM TM}
\right], \\ H_t &=& \sum_{<f,g>,\sigma} \Bigl\{ t_{fg}^{SS} X_{f}^{S
\bar{\sigma}} X_{g}^{\bar{\sigma} S}\\ &+& t_{fg}^{TT} \left( \sigma
\sqrt{2} X_{f}^{T0 \bar{\sigma}} - X_{f}^{T2\sigma \sigma} \right) \left(
\sigma \sqrt{2} X_{g}^{\bar{\sigma} T0} - X_{g}^{\sigma T2\sigma} \right)\\
&+& t_{fg}^{ST} 2 \sigma \gamma_{b} \left[ X_{f}^{S \bar{\sigma}} \left(
\sigma \sqrt{2} X_{g}^{\bar{\sigma} T0} - X_{g}^{\sigma T2\sigma} \right)
+H.c. \right] \Bigl\}.
\end {eqnarray*}
Upper indexes of hopping integrals ($0$,$S$,$T$) corresponds to excitations
which accompanied by hopping from site $f$ to $g$, i.e. in Hamiltonian one
have the following terms: $\sum\limits_{<f,g>,\sigma} t_{fg}^{MN}
X_f^{M \sigma} X_g^{\sigma N}$. The relation between these effective
hoppings and microscopic parameters of multiband $p-d$ model is as follows:
\begin{eqnarray}
t_{fg}^{00} &=& -2t_{pd}\mu_{fg}2uv - 2t_{pp}\nu_{fg}v^2, \nonumber \\
t_{fg}^{SS} &=& -2t_{pd}\mu_{fg}2\gamma_x\gamma_b - 2t_{pp}\nu_{fg}\gamma_b^2, \nonumber \\ t_{fg}^{0S} &=& -2t_{pd}\mu_{fg}(v\gamma_x+u\gamma_b) - 2t_{pp}\nu_{fg}v\gamma_b, \label{eq:t} \\ t_{fg}^{TT} &=& \frac{2t_{pd}}{\sqrt{3}}\lambda_{fg}2\gamma_a\gamma_z +2t_{pp}\nu_{fg}\gamma_a^2 - 2t'_{pp}\lambda_{fg}2\gamma_p\gamma_a, \nonumber \\ 
t_{fg}^{ST} &=& \frac{2t_{pd}}{\sqrt{3}}\xi_{fg}\gamma_z + 2t_{pp}\chi_{fg}\gamma_a - 2t'_{pp}\xi_{fg}\gamma_p, \nonumber
\end{eqnarray}
The factors $\mu$, $\nu$, $\lambda$, $\xi$, $\chi$ are the coefficients of
Wannier transformation made in GTB method and $u$, $v$, $\gamma_a$,
$\gamma_b$, $\gamma_z$, $\gamma_p$ are the matrix elements of annihilation-creation operators in the Hubbard X-operators representation.

The resulting Hamiltonian (\ref {eq:Heff}) is the generalization of the
$t-J$ model to account for two-particle triplet state. Significant feature
of effective singlet-triplet model is the asymmetry for n- and p-type
systems which is known experimentally. So, we can conclude that for n-type
systems the usual $t-J$ model takes place while for p-type superconductors
with complicated structure on the top of the valence band the
singlet-triplet transitions plays an important role. 

Using set of microscopic parameters (\ref {eq:pparams}) in the Table~\ref{table0} we present 
numerical values of the hopping and exchange parameters calculated according to (\ref{eq:t}).

\begin {table}
\caption {\label {table0} Parameters of the effective singlet-triplet model for p-type cuprates obtained in the framework of GTB method (all values in eV).}
\begin {ruledtabular}
\begin {tabular}{c|llllll}
$\rho$ &$t^{00}_{\rho}$ &$t^{SS}_{\rho}$ &$t^{0S}_{\rho}$ &$t^{TT}_{\rho}$ &$t^{ST}_{\rho}$ &$J_{\rho}$ \\
\hline
(0,1)  & 0.373 &  0.587 & -0.479 & 0.034  & 0.156 & 0.115   \\
(1,1)  & 0.002 & -0.050 &  0.015 & -0.011 & 0     & 0.0001	\\
(0,2)  & 0.050 &  0.090 & -0.068 & 0.015  & 0.033 & 0.0023 \\
(2,1)  & 0.007 &  0.001 & -0.006 & -0.004 & 0.001 & 0 \\
\end {tabular}
\end {ruledtabular}
\end {table}

\section {Comparison of parameters}

The resulting parameters from {\it ab initio} \cite {Sorella} and GTB 
calculations  are presented in Table~\ref {table1}. Here $\rho$ 
is the connecting vector between two copper centers, $t_{\rho}$ is 
the hopping parameter (in the effective singlet-triplet model it is 
equal  to $t_{\rho}^{SS}$, see (\ref {eq:t})), $J_{\rho}$ is the 
antiferromagnetic exchange integral.

\begin {table}
\caption {\label {table1}Comparison of {\it ab initio} parameters \cite {Sorella} and 
parameters, obtained in the framework of GTB method (all values in eV).}
\begin {ruledtabular}
\begin {tabular}{c|ll|ll}
&  \multicolumn {2}{c|}{{\it ab initio}} & \multicolumn {2}{c}{GTB method} \\
$\rho$ & $t_{\rho}$ & $J_{\rho}$ & $t_{\rho}$ & $J_{\rho}$ \\
\hline
(0,1)  &  0.486     & 0.109      &  0.587     & 0.115   \\
(1,1)  & -0.086     & 0.016      & -0.050     & 0.0001	\\
(0,2)  & -0.006     & 0          &  0.090     & 0.0023 \\
(2,1)  &  0         & 0          &  0.001     & 0 \\
\end {tabular}
\end {ruledtabular}
\end {table}

As one can see, despite slight differences, parameters of both methods are
very close  and show similar dependence on distance. It is worth to
mention that both methods give disproportionality between $t_{\rho}$ and
$J_{\rho}$. In the usual $t-J$ model proportionality ($J_{\rho}=2
t_{\rho}^2 / U$) takes place as soon as this $t-J$ model obtained from the
Hubbard model with Hubbard repulsion $U$. In case of the singlet-triplet
model the intersubband hoppings $t_{\rho}^{0S}$ which determines value of
$J_{\rho}$ is different (\ref {eq:t}) from the intrasubband hoppings
$t_{\rho}^{SS}$ which determines $t_{\rho}$. This fact resulted in
more complicated relation between $t_{\rho}$ and $J_{\rho}$.

In the framework of LDA band structure of $YBa_2CuO_{7+x}$ and withing
orbital projection approach it was shown \cite {Andersen} that the 1-band
Hamiltonian reduced from eight-band Hamiltonian should include not only the
nearest-neighbor hopping terms ($t$), but also second ($t'$) and third
($t''$) nearest-neighbors hoppings. In GTB method the dependence of
hoppings $t_{\rho}$ on distance automatically results from the distance
dependence of coefficients of Wannier transformation made in this method
(see Eq.~(\ref {eq:t})). In order to show correspondence between results of
various authors, we present comparison of our parameters and parameters,
widely used by different groups in Table~\ref {table2}.

\begin{table*}
\caption{\label{table2}Comparison of calculated parameters and parameters, used in the literature.}
\begin{ruledtabular}
\begin{tabular}{l|llllllllllll}
&0\footnotemark[1] &0\footnotemark[2] &I\footnotemark[3] &II\footnotemark[3] &III\footnotemark[3] 
&IV\footnotemark[3] &V\footnotemark[3] &VI\footnotemark[3] &VII\footnotemark[4] 
&VIII\footnotemark[4] &IX\footnotemark[5] &X\footnotemark[5]\\
&LSCO&LSCO&LSCO&LSCO&LSCO&Bi2212&YBCO&SCOC&YBCO&LSCO&LSCO&YBCO\\
quantity &here &here &\cite{Nazarenko} &\cite{Belinicher1,Belinicher2} 
&\cite{Tohayama1,Tohayama2} &\cite{Tohayama1,Tohayama2,Kim} 
&\cite{Onufrieva} &\cite{Eder} &\cite{Andersen} &\cite{Pavarini} &\cite{Brenig} &\cite{Brenig}\\
\hline
$t$, eV & 0.587 &0.486 & 0.416 & 0.35 & 0.35 & 0.35 & 0.40 & 0.40 & 0.349 & 0.43 & -- & -- \\
$t'/t$    &-0.085 &-0.18 &-0.350 &-0.20 &-0.12 &-0.34 &-0.42 &-0.35 &-0.028 &-0.17 & -- & -- \\
$t''/t$   & 0.154 &0.012 & --    & 0.15 & 0.08 & 0.23 &-0.25 & 0.25 & 0.178 & --   & -- & -- \\
$J$, eV & 0.115 &0.109 & 0.125 & 0.14 & 0.14 & 0.14 & 0.17 & 0.12 & --    & --   & 0.126 & 0.125, 0.150\\
$J/|t|$     & 0.196 &0.224 & 0.300 & 0.40 & 0.40 & 0.40 & 0.43 & 0.30 & --    & --   & -- & --
\end{tabular}
\end{ruledtabular}
\footnotetext[1]{GTB method parameters}
\footnotetext[2]{{\it ab initio} parameters obtained in present paper}
\footnotetext[3]{parameters obtained by fitting to experimental data}
\footnotetext[4]{{\it ab initio} parameters}
\footnotetext[5]{parameters obtained from two-magnon Raman scattering}
\end{table*}

The parameters extracted from experimental data are listed in columns I-VI
of Table~\ref {table2}. LDA calculated parameters are presented in
columns VII and VIII. Our results for hoppings are in the best agreement
with columns III, VII and VIII. This similarity is not surprising. In LDA
calculations the bandwidth of strongly correlated electron systems are
usually overestimated due to the lack of proper account of strong Coulomb
repulsion of electrons. But it is well known that the Fermi surface
obtained by this method is in very good agreement with experiments. The
main contribution to the shape of the Fermi surface comes from kinetic energy of the electrons (hopping parameters), so values of hoppings should be properly estimated by LDA calculations (columns VII, VIII). In works by Tohayama and Maekawa \cite {Tohayama1,Tohayama2} (column III) the parameters was obtained by fitting the LSCO tight-binding Fermi
surface to the experimental one. This procedure should give the same values
as an LDA calculations and, as one can see, it does. By the same technique
the parameters for $Bi_2Sr_2CaCu_2O_{8+x}$ (Bi2212, column IV) was obtained
\cite {Tohayama1,Tohayama2}. These parameters are different from LSCO case
and present paper and the most straightforward explanation is the more
complicated structure of the Fermi surface of Bi2212 compound. In the
present paper the single-layer (LSCO-like) compounds are considered and
effects of multiple $CuO_2$-planes (i.e. bi-layer splitting) are
neglected.  The difference between our hoppings and hoppings in column V
appeared due to the same reason (in Ref.~\cite {Onufrieva} the
$YBa_2Cu_3O_6$ insulating compound was investigated).

In the last two columns of Table~\ref {table2} the antiferromagnetic
exchange parameter $J$ obtained from two-magnon Raman scattering analysis
by moment expansion (LSCO, column IX) and spin-wave theory (YBCO, column X)
are presented (for details see review~\cite {Brenig} and references
therein). Our values of $J$ (column 0) are in good agreement
with values extracted from experiments and similar to listed in columns
I-VI.

In the paper \cite {Katanin} the Heisenberg Hamiltonian on the square
lattice with plaquette ring exchange was investigated. The fitted exchange
interactions $J=0.151$~eV, $J'=J''=0.025 J$ gave values for the spin
stiffness and the Neel temperature in excellent agreement with experimental
data for insulating compound $La_2CuO_4$. In GTB calculations $J=0.115$~eV,
$J'=0.0009 J$ and $J''=0.034 J$. The values of $J$ are close to each other
but different. This difference could be explained by the fact that authors
of \cite {Katanin} used the Heisenberg Hamiltonian and inclusion of hopping term
should renormalize presented values of exchange interactions. Agreement
between $J''$ in GTB calculations and Ref.~\cite {Katanin} is good but
values of $J'$ is completely different. The last issue could be addressed
to oversimplification of calculations in \cite {Katanin} -- the authors put
$J'=J''$ by hand to restrict number of fitting parameters.

\begin{figure}
\includegraphics[width=\linewidth]{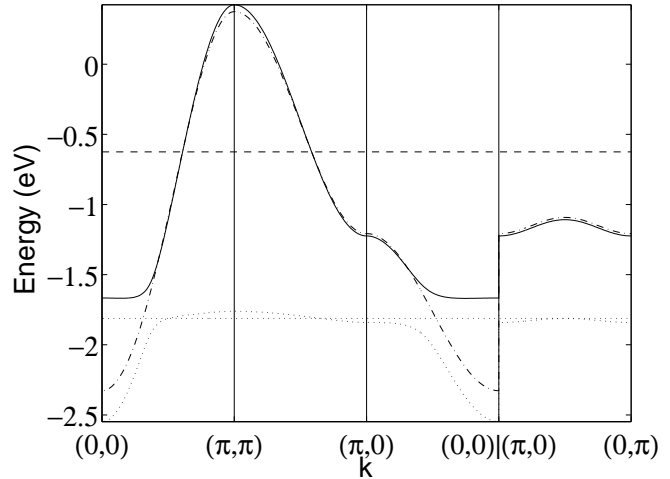}
\caption{\label{fig3}
Dispersion curves on top of the valence band for effective singlet-triplet model
(singlet subband - solid line, triplet subbands - dotted lines) and $t-t'-J$ model (dash-dotted line) at optimal doping $x=0.15$,
dashed line denotes self-consistently obtained chemical potential $\mu$.}
\end{figure}

Now we will discuss the difference between our parameters and parameters in
columns I,II,VI and column IV (SCOC). The hoppings in mentioned papers were
obtained by fitting $t-t'-t''-J$ model dispersion to the experimental ARPES
spectra \cite {Wells,Kim} for insulating $Sr_2CuO_2Cl_2$. We claim that the
discrepancy between GTB method results and $t-t'-t''-J$ model results stems
from absence of singlet-triplet hybridization in latter model. This
statement can be proved by comparison of ''bare'' $t-t'-J$ model
(\ref {eq:HtJ}) dispersion and dispersion of singlet-triplet $t-t'-J$ model
(\ref {eq:Heff}). The paramagnetic non-superconductive phase was
investigated in Hubbard-I approximation both in the singlet-triplet and
$t-t'-J$ models. Results for optimal doping (concentration of holes
$x=0.15$) are presented in Fig.~\ref{fig3}.

There is strong mixture of singlet and triplet bands along $(0,0)-(\pi,\pi)$ and $(\pi,0)-(0,0)$ 
directions due to the $t^{ST}$ matrix element (see (\ref{eq:t})) in both paramegnetic (Fig.~\ref{fig3}) 
and in antiferromagnetic phases (Fig.~\ref{fig1}). It is exactly the admixture of 
the triplet states that determines in our approach the dispersion and the ARPES data in the undoped 
SCOC at the energies $0.3\div0.4$~eV below the top of the valence band, where the $t-t'-J$ 
model \cite{Nazarenko} failed, and the additional $t''$ parameter have been involved in 
the $t-t'-t''-J$ model \cite{Belinicher1,Tohayama1}.
In our approach this parameter is not as necessary as in ''bare'' $t-t'-J$ model
since the singlet-triplet hybridization is included explicitly.

In Ref.~\cite {Xiang} the $t-t'-t''-J$ model was also used to describe
dispersion of insulating $Sr_2CuO_2Cl_2$ and the set of parameters was the
same as in Refs.~\cite {Tohayama1,Tohayama2}. But the authors of
Ref.~\cite {Xiang} used completely different definition of hopping
parameters: in their paper the $t'$ term stands for hopping between
two nearest neighboring oxygens and the $t''$ term stands for the hopping
between two oxygens on the two sides of Cu. Such definition is completely
different from used in other cited papers where $t$, $t'$, $t''$ terms
stands for hoppings between plaquettes centered on copper sides and we
can't make comparison with their results.

The analysis of Table~\ref{table2} data gives the following ranges for different
parameters: $0.350\div0.587$~eV for $t$, $-0.420\div-0.028$ for $t'/t$,
$0.012\div0.250$ for $t''/t$ with the exception of value in Ref.~\cite{Onufrieva}
and $0.115\div0.150$~eV for $J$. In general one can see close similarity in the first neighbor hopping $t$ and interaction $J$ for the different methods and materials, and more discrepancy in such subtle parameters as $t'$ and $t''$.

\section {Conclusion}

One of the significant results of this paper is the relationship (\ref{eq:t}) between 
microscopic parameters and parameters of the effective singlet-triplet model. Thus 
the effective model parameters are not free any more and have a direct physical
meaning coming from the dependence on microscopic parameters. The parameters of
the effective singlet-triplet model were obtained both from {\it ab initio} and
model calculations. Model calculations were performed in the framework of GTB
method for insulating single-layer copper oxide superconductor.  
The {\it ab initio} calculations for $La_2CuO_4$ were done by conventional 
LDA TB-LMTO method. The agreement between parameters is remarkably good. 
Obtained parameters are also in good agreement with widely used parameters 
of the $t-t'-t''-J$ model but some difference exists. This difference is 
attributed to the neglect of triplet excitations in simple $t-t'-t''-J$ model.  
After careful analysis we proposed the set of parameters for effective models 
(e.g. $t-t'-t''-J$ model or effective singlet-triplet model)
for proper quantitative description of physics of hole doped High-$T_c$ cuprates.

\begin {acknowledgments} M.M.K., V.A.G., and S.G.O. thank the Free
University of Berlin for hospitality during  their stay.  This work has
been supported by INTAS grant 01-0654, Joint Integration Program of
Siberian and Ural Branches of Russian Academy of Science, Russian
Foundation for Basic Research grant 03-02-16124, Russian Federal Program
``Integratsia'' grant B0017, Program of the Russian Academy of Science
``Quantum Macrophysics'', and Siberian Branch of Russian Academy of Science
(Lavrent'yev Contest for Youth Projects),  RFFI-01-02-17063 (VIA, IAN,
MAK), RFFI (MAS)-03-02-06126 (IAN), Ural Branch of Russian Academy  of
Science for Young Scientists (IAN, ZVP), Grant of the President of Russia 
MK-95.2003.02 (IAN).   
\end {acknowledgments}


\begin {thebibliography}{9}
\bibitem {qqq1} Z.-X. Shen and D.S. Dessau, Phys. Rep. 253, 1 (1995); 
								E. Dagotto, Rev. Mod. Phys. 66, 763 (1994); 
								A.P. Kampf, Phys. Rep. 249, 219 (1994)
%
\bibitem {Emery} V.J. Emery, Phys. Rev. Lett. 58, 2794 (1987)
\bibitem {Varma} C.M. Varma et al., Solid State Commun. 62, 681 (1987)
\bibitem {Gaididei} Yu. Gaididei and V. Loktev, Phys. Status Solidi B 147, 307 (1988)
%
\bibitem {lmto} O.K. Andersen and O. Jepsen, Phys. Rev. Lett. 53, 2571 (1984) 
\bibitem {tblmto} O.K. Andersen, Z. Pawlowska and O. Jepsen, Phys. Rev. B 34, 5253 (1986) 
\bibitem {struc} J.D. Axe and M.K. Crawford, J. Low Temp. Phys. 95, 271 (1994)
\bibitem {Sorella} V. I. Anisimov et al., Phys. Rev. 66, 100502 (2002)
\bibitem {exch} A.I. Lichtenstein et al., J. Magn. Mag. Matter 67, 65 (1987); 
								A.I. Lichtenstein, V.I. Anisimov and J. Zaanen, Phys. Rev. B 52, R5467 (1995)
\bibitem {lda_u} V.I. Anisimov, J. Zaanen and O. Andersen, Phys. Rev. B 44, 943 (1991); 
								 V.I. Anisimov et al., J. Phys.: Condens. Matter 9, 767 (1997) 
\bibitem {constrain} O. Gunnarsson et al., Phys. Rev. B 39, 1708 (1989); 
										 V.I. Anisimov and O. Gunnarsson, $ibid.$ 43, 7570 (1991)
%
\bibitem {Chao} K.A. Chao, J. Spalek and A.M. Oles, J. Phys. C: Sol. Stat. Phys. 10, 271 (1977)
\bibitem {Hubbard} J.C. Hubbard, Proc. Roy. Soc. A 276, 238 (1963)
\bibitem {Hybertsen} M.S. Hybertsen, M. Schluter and N.E. Christensen, Phys. Rev. B 39, 9028 (1989)
\bibitem {McMahan} A.K. McMahan, J.F. Annett and R.M. Martin, Phys. Rev. B 42, 6268 (1990)
%
\bibitem {Bianconi} A. Bianconi et al., Phys. Rev. B 38, 7196 (1988);
\bibitem {Romberg} H. Romberg et al., Phys. Rev. B 41, 2609 (1990)
\bibitem {Chen} C.H. Chen et al., Phys. Rev. Lett. 68, 2543 (1992)
%
\bibitem {Ovchinnikov} S.G. Ovchinnicov and I.S. Sandalov, Physica C 161, 607 (1989)
\bibitem {Gavrichkov1} V.A. Gavrichkov et al., Phys. Rev. B 64, 235124 (2001)
\bibitem {Gavrichkov2} V.A. Gavrichkov et al., Zh. Eksp. Teor. Fiz. 118, 422 (2000); [JETP 91, 369 (2000)]
%
\bibitem {Wells} B.O. Wells et al., Phys. Rev. Lett. 74, 964 (1995)
\bibitem {Durr} C. D{\"u}rr et al., Phys. Rev. B 63, 014505 (2000)
%
\bibitem {Borisov1} A.A. Borisov, V.A. Gavrichkov and S.G. Ovchinnikov, Mod. Phys. Lett. B 17, 479 (2003)
\bibitem {Borisov2} A.A. Borisov, V.A. Gavrichkov and S.G. Ovchinnikov, Zh. Eksp. Teor. Fiz. 124, 862 (2003); [JETP 97, 773 (2003)]
%
\bibitem {Harima} N.Harima et al, Phys. Rev B 64, 220507(R) (2001)
\bibitem {Ino1} A. Ino et al., Phys. Rev. Lett. 79, 2101 (1997)
\bibitem {Ino2} A. Ino et al., Phys. Rev. B 65, 094504 (2002)
%
\bibitem {Zaanen} J. Zaanen, A.M. Oles, P. Horsch, Phys. Rev. B 46, 5798 (1992)
\bibitem {Hayn} R. Hayn et al., Phys. Rev. B 47, 5253 (1993)
%
\bibitem {Korshunov} M. Korshunov and S. Ovchinnikov, Fiz. Tv. Tela 43, 399 (2001) [Phys. Sol. State 43, 416 (2001)]
\bibitem {Raimondi} R. Raimondi et al., Phys. Rev. B 53, 8774 (1996)
%
\bibitem {Andersen} O.K. Andersen et al., J. Phys. Chem. Solids 56, 1573 (1995)
\bibitem {Nazarenko} A. Nazarenko et al., Phys. Rev. B 51, 8676 (1995)
\bibitem {Belinicher1} V.I. Belinicher, A.I. Chernyshev and V.A. Shubin, Phys. Rev. B 53, 335 (1996)
\bibitem {Belinicher2} V.I. Belinicher, A.I. Chernyshev and V.A. Shubin, Phys. Rev. B 54, 14914 (1996)
\bibitem {Tohayama1} T. Tohayama and S. Maekawa, Supercond. Sci. Technol. 13, R17 (2000)
\bibitem {Tohayama2} T. Tohayama and S. Maekawa, Phys. Rev. B 67, 092509 (2003)
\bibitem {Kim} C. Kim et al., Phys.Rev. Lett. 80, 4245 (1998)
\bibitem {Onufrieva} F.P. Onufrieva, V.P. Kushnir and B.P. Toperverg, Phys. Rev. B 50, 12935 (1994)
\bibitem {Eder} R. Eder, Y. Ohta, G.A. Sawatzky, Phys. Rev. B 55, R3414 (1997)
\bibitem {Pavarini} E. Pavarini et al., Phys. Rev. Lett. 87, 047003 (2001)
\bibitem {Brenig} W. Brenig, Phys. Rep. 251, 153 (1995)
%
\bibitem {Katanin} A.A. Katanin and A.P. Kampf, Phys. Rev. B 66, 100403(R) (2003)
\bibitem {Xiang} T. Xiang and J.M. Wheatley, Phys. Rev. B 54, R12653 (1996)
\end {thebibliography}
\end {document}